\documentclass[twocolumn]{aastex631}
\usepackage{appendix}

\shorttitle{super-Earth formation with ohmic dissipation}
\shortauthors{Jia et al.}

\begin{document}

\title{Ohmic dissipation during the formation of super-Earth}

\correspondingauthor{Shi Jia}
\email{sjia@must.edu.mo}

\correspondingauthor{Cong Yu}
\email{yucong@mail.sysu.edu.cn}

\author{Shi Jia}
\affiliation{State Key Laboratory of Lunar and Planetary Sciences, Macau University of Science and Technology, Macau, People’s Republic of China}
\affiliation{CNSA Macau Center for Space Exploration and Science, Macau, People's Republic of China}

\author{Wei, Zhong}
\affiliation{School of Physics and Astronomy, Sun Yat-Sen University, Zhuhai, 519082, People’s Republic of China}
\affiliation{State Key Laboratory of Lunar and Planetary Sciences, Macau University of Science and Technology, Macau, People’s Republic of China}
\affiliation{CSST Science Center for the Guangdong-Hong Kong-Macau Greater Bay Area, Zhuhai, 519082, People’s Republic of China}

\author{Cong Yu}
\affiliation{School of Physics and Astronomy, Sun Yat-Sen University, Zhuhai, 519082, People’s Republic of China}
\affiliation{State Key Laboratory of Lunar and Planetary Sciences, Macau University of Science and Technology, Macau, People’s Republic of China}
\affiliation{CSST Science Center for the Guangdong-Hong Kong-Macau Greater Bay Area, Zhuhai, 519082, People’s Republic of China}

\begin{abstract}
Super-Earth population, as one of the representatives of exoplanets, plays an important role in constraining the planet formation theories. According to the prediction from core-accretion models, super-Earths should be rare because their masses are in the range of the critical mass above which they would grow to be gas giants by runaway gas accretion. In this work, we investigate the effect of ohmic dissipation on the planetary thermal structure and cooling contraction as planets accrete gas from their surrounding disks. We find that the extra heating energy from Ohmic heating deposited into planetary envelopes can push the planetary radiative-convective boundaries inward and prevent the planets from cooling, and can even halt accretion. We explore 
parameter space to study the dependence of cooling timescale on the input parameters of the ohmic-dissipation model. 
Numerical results show that gas accretion can be halted before runaway gas accretion and the envelope mass is only several percent of planetary core mass for some parameter sets.
Our results suggest that ohmic dissipation is a potential mechanism to delay the gas accretion and promote the formation of super-Earths. Future observations may help to constrain the importance of ohmic dissipation on the super-Earth formation.

\end{abstract}


\section{Introduction} \label{intro}

Super-Earths (or sub-Neptunes), generally defined as the planet mass (or radius) between that of Earth and Neptune, contribute a significant part of the current exoplanet population. These low mass planets commonly orbit around Sun-like stars with orbital periods shorter than 100 days (e.g., Howard et al. 2010; Batalha et al. 2013; Petigura et al. 2013; Zhu et al. 2018). Moreover, most of super-Earths possess low bulk densities, implying that gas envelopes should be held outside their solid cores (e.g., Weiss \& Marcy 2014; Wu \& Lithwick 2013b; Rogers 2015). Detailed planetary interior models suggest that the mass ratios of the outer envelope and the inner core are typically from  several up to about ten percent(e.g., Lopez \& Fortney 2014; Wolfgang \& Lopez 2015). 

Currently, the formation of super-Earths is still an open question. A lot of studies have studied their formation from several conceivable angles. For example,
some concentrated on exploring the conditions for the in situ formation of super-Earths. Lee et al.(2014) found that super-Earths can be formed in dusty and gas-poor disks (see also Lee \& Chiang 2015 and Lee et al. 2018).
Others focused on studying the possible mechanisms that can promote the formation of super-Earths. For example, giant impact(e.g., Inamdar \& Schlichting 2015; Liu et al. 2015) and tidal heating (Ginzburg \& Sari 2017) could supply extra heating energy to delay or halt the contraction of protoplanets. Other factors, such as the tidally forced turbulence (Yu 2017) and planetary rotation (Zhong \& Yu 2021), could change the planet's interior structure, especially for the location of radiative-convective boundary, and slow down the envelope contraction. Furthermore, the atmospheric recycling, which is also referred to entropy advection (Ali-Dib et al. 2020), is also important to prevent the super-Earths from experiencing runaway accretion and growing to be gas giants (e.g., Ormel et al. 2015a; Ormel et al. 2015b; B{\'e}thune \& Rafikov 2019a; B{\'e}thune \& Rafikov 2019b).   

In this work, we explore the effect of ohmic dissipation on the evolution of planetary envelope as protoplanets accrete gas from their natal disks.
Ohmic dissipation has been widely used to study the anomalous radius of Hot Jupiters (e.g., Batygin \& Stevenson 2010; Perna et al. 2010; Batygin et al. 2011; Huang \& Cumming 2012; Wu \& Lithwick 2013a). Pu \& Valencia (2017) also investigate the radius evolution of mini-Neptunes with Ohmic heating. 
The ohmic dissipation of close-in planets can be driven by the thermal dynamics (e.g., Showman et al. 2010; Menou 2012). Due to the strong irradiation from host stars, the surface temperature of close-in planets can be higher than 1000 K, resulting in the ionization of alkali metals. A thermal wind in the planetary upper atmosphere is caused by the temperature difference in the atmosphere between the day side and night side because of the tidal locking of close-in planet (e.g., Showman et al. 2010; Menou 2012). With the magnetic field from the close-in planet itself or its host star, induced current in planetary envelope is then generated and penetrates into planet's interior.
As a result, heating power from ohmic dissipation is deposited into the planetary envelope. In addition, the Ohmic dissipation of close-in planets can also be motivated by the magnetic interaction with their host stars (e.g., Laine \& Lin 2008; Laine \& Lin 2012). 

It is possible that ohmic dissipation is induced inside the planetary envelope during planetary gas accretion phase. Although the protoplanets can not obtain irradiation energy directly from their host stars because they are embedded into the disks, the close-in protoplanets can still have high surface temperatures. According to the minimum-mass extrasolar nebula model of Chiang \& Laughlin (2013), the disk temperature at the location with a separation 0.1 AU from host star is on the order of 1000 K, which means that the gas material with solar-like metallicity at this location can be weakly ionized. The weakly ionized material is accreted onto the surface of planets and the angular momentum of accreting gas flow is then transported inward to spin up the planets (e.g., Dittmann 2021). Atmospheric wind is formed in the planetary envelope during the accretion process. Therefore, the close-in protoplanet will undergo the ohmic dissipation while the weakly ionized accreting gas flows across the planetary intrinsic magnetic fields or the fields from their host stars.

This work is organized as follows. In Section \ref{model}, we describe the methodology that we used to model the planetary gas accretion with ohmic dissipation and give the basic parameters and boundary conditions. We then present the numerical results of planetary thermal structures and cooling contraction in Section \ref{resul}. Finally, we discuss the model uncertainties and give a summary in Section \ref{concl}.

\section{Model and method} \label{model}
Coupling the ohmic heating within the planet interior structure to investigate the gas-accretion process is complex. In the first step, we adopt an simplified Ohmic heating model from Wu \& Lithwick (2013a) and Pu \& Valencia (2017) to construct the planet's evolution model. A detailed analysis of Ohmic dissipation is beyond this scope.
In this section, we first review the Ohmic heating mechanism and planet's interior model. We then describe our method to model the gas-accretion coupling with the ohmic heating mechanism. 

\subsection{Ohmic dissipation}\label{ohmic}

The intensity of ohmic dissipation in the planetary envelope is determined by several factors, including the magnetic field acting on the planet, wind speed and wind zone depth in the envelope, and electrical conductivity of the material in the envelope. We assume that the planet has a  magnetic dipole field and wind zone with constant wind velocity (Wu \& Lithwick 2013a). Ohmic dissipation in our model is constrained between the outer envelope surface and the inner core, which is treated as an insulator (e.g., Pu \& Valencia 2017). 
 
\subsubsection{Electrical conductivity}  
The electrical conductivity of planetary envelope comes from the contribution of alkali metal ionization in the upper envelope and hydrogen ionization in the deeper envelope. We adopt the conductivity profile from Pu \& Valencia (2017).
The element potassium (K) has the lowest ionization potential for the alkali metal. The conductivity of alkali metal is dominated by the element K at low pressures. The conductivity of element K is given by (Pu \& Valencia 2017):
\begin{equation}
     \sigma_{Z} \simeq  1.74 \times 10^3 \left( \frac{T}{1600 \ K} \right)^{3/4} \left( \frac{P}{bar} \right)^{(-1/2)} \mathrm{exp\left(\frac{-4.35 eV}{2k_BT}\right)}
\end{equation}
Only considering the contribution of element K for the conductivity of alkali metal is appropriate to our situation. Compared with the full implementation, this approximation is accurate to within a few percent below the pressures of $\sim$ 100 bars (Pu \& Valencia 2017).

In the deeper part of planetary envelope, hydrogen is ionized by high pressure. The conductivity of the hydrogen is (e.g., Liu et al. 2008; Huang \& Cumming 2012) expressed as follows:
\begin{equation}
    \sigma_{X}=\sigma_0 \ \mathrm{exp\left(\frac{-E_g(\rho)}{2k_B T} \right)},
\end{equation}
where $\mathrm{E_g=20.3-64.7 \rho}$  is in eV,  $\rho$ is the density in mol $cm^{-3}$, and $\sigma_0=3.4 \times 10^{20} \ s^{-1} \ \mathrm{exp(-44 \rho)} $  .

The total conductivity is then given as 
\begin{equation}
    \sigma = \sigma_X + \sigma_Z
\end{equation}

\subsubsection{Ohmic heating model} 

As the weakly ionized atmospheric flow moves across the intrinsic magnetic field of the planet or the field from its host star,  an induced current is generated in the planetary envelope. Simultaneously, an induced magnetic field is produced. 
In the steady state, the induction equation is expressed as:
\begin{equation}
  \frac{\partial \textbf{B} }{\partial t} = \nabla \times (\textbf{v} \times \textbf{B}) - \nabla \times \eta (\nabla \times \textbf{B}) = 0 
\end{equation}
where $\eta$  is the magnetic diffusivity, $\eta = c^2/ 4 \pi \sigma $. 

The magnetic field acting on the planet is assumed to be a dipolar field (Wu \& Lithwick 2013a): 
\begin{equation}
   \mathrm{ \textbf{B}_{dip}=M_B\frac{2\cos\theta \textbf{e}_r + \sin\theta \textbf{e}_\theta}{r^3} }
\end{equation}
where $M_B$ is the magnetic dipole moment of the planet, $M_B = \mathrm{B R^3}$. $\mathrm{B}$ is the surface dipole field strength. $\mathrm{R}$ is planetary radius.

Assuming a constant angular velocity $\omega$ in the wind zone, $\textbf{v}$ is given by:
\begin{equation}
    \textbf{v} = \omega r \ \sin \theta  \textbf{e}_\phi
\end{equation}
In the wind zone, the current is given with Ohm's law:
\begin{equation} \label{jw}
   \mathrm{\textbf{J}=\sigma \left( - \nabla \Phi + \frac{\textbf{v} \times \textbf{B}_{dip}}{c}  \right)}
\end{equation}
Where $\Phi$ is the electric potential and $\textbf{v}$ is the wind velocity. Below the wind zone, the current is obtained by:
\begin{equation} \label{jnw}
   \mathrm{\textbf{J}=- \sigma \nabla \Phi}
\end{equation}
With $\nabla \cdot \textbf{J} = 0$, Equations (\ref{jw}) and (\ref{jnw}) are solved numerically (details see Appendix \ref{appexa}).

The ohmic power can be obtained with the electric current profile: 
\begin{equation}
    P_{\mathrm{oh}}= \int \frac{J^2}{\sigma} dV
\end{equation}
where $J=\sqrt{J_r^2 + J_\theta^2}$, $J_r$ and $J_\theta$ are the radial current and the meridional current, respectively. The Ohmic power profile along the radius is expressed as:  
\begin{equation}\label{eq:dpdr}
    \frac{d P_{\mathrm{oh}}}{dr} = 4 \pi r^{2}  \frac{J^2}{\sigma}
\end{equation}

The boundary conditions for the radial current $J_r$ are that $J_r=0$ at the planetary surface $R_\mathrm{out}$ and at the core surface $R_c$. 

\subsection{Planet model}
We adopt the planet interior structure similar to the model from Piso \& Youdin(2014). We employ the realistic equation of state (EOS) for the hydrogen-helium mixture and dust-free opacity\footnote{We adopt the dust free opacity as our fiducial case since the dust grains could coagulate efficiently (e.g., Ormel 2014; Mordasini et al. 2014). For the location of protoplanets we considered here, a=0.1 AU, the coagulation timescale is several orders of magnitude smaller than the disk lifetime (e.g., Lee et al. 2014). We also briefly discuss the envelope accretion with dust opacity in Section \ref{topa}.}, details are shown below. 

The basic structure equations are as follows:
\begin{equation}
    \frac{dm}{dr} = 4 \pi \rho r^2
\end{equation}

\begin{equation}
    \frac{dP}{dr} =  - \rho \frac{Gm}{r^2}
\end{equation}

\begin{equation}
    \frac{dT}{dr} =\nabla \frac{T}{P} \frac{dP}{dr}
\end{equation}
where $r$ is radius, $m$ is the mass enclosed within the radius $r$, $P$ is pressure, $T$ is temperature, $\rho$ is density, $G$ is gravitational constant. The temperature gradient, $\nabla = d \ ln T/ d \ ln P$, is set to the minimum value of radiative gradient $\nabla_{rad}$ and adiabatic gradient $\nabla_{ad}$ :     
\begin{equation}
    \nabla = \min(\nabla_{ad}, \nabla_{rad}),
\end{equation}
\begin{equation}
    \nabla_{rad} = \frac{3 \kappa P L}{16 \pi \sigma G M T^4}
\end{equation}
\begin{equation}
    \nabla_{ad} = \left( \frac{d \ln T }{d \ln P} \right)_{ad}
\end{equation}
where $L$ is the luminosity, $\kappa$ is opacity, and $\sigma$ is the Stefan–Boltzmann constant. 

The ohmic heating model is coupled into the planet structure model. With the extra energy source from the ohmic heating, the total luminosity is now expressed as:
\begin{equation}
    L = L_{cool} + P_{oh}
\end{equation}
where $L_{cool}$ is the cooling luminosity from the planet convective interior. $L_{cool}$ is assumed spatially constant inside of planet (e.g., Piso \& Youdin 2014; Lee et al. 2014). The planetary luminosity profile is obtained with Equation (\ref{eq:dpdr}):
\begin{equation}
    \frac{dL}{dr} =  \frac{d P_{\mathrm{oh}}}{dr}
\end{equation}

The EOS for hydrogen-helium mixture of the envelope used here is from the EOS table of Chabrier et al.(2019; 2021). The metallicity of EOS that we used is Z=1-X-Y=0.017 (X=0.708,Y=0.275). The opacity is from the opacity table of Freedman et al.(2014). The EOS and opacity are solar-like.

The cooling time is given by (Piso \& Youdin 2014):
\begin{equation} \label{time}
    \Delta t = \frac{-\Delta E + \langle e \rangle \Delta M - \langle P \rangle \Delta V_{\langle M \rangle} }{\langle L_{\mathrm{cool}} \rangle}
\end{equation}
where $\Delta$ denotes the difference between the two adjacent snapshots and the brackets indicate their average. The $e$ in second term of Equation (\ref{time}) is the specific energy of the accreting matter: $e=-Gm/r +u$. The third term in Equation (\ref{time}) is the work done by the envelope contraction. The total energy is given as follows:
\begin{equation} \label{energy}
    E = \int_{M_c}^{M} u \ dm - \int_{M_c}^{M} \frac{G m}{r}dm 
\end{equation}
where $u$ is the specific internal energy, which is interpolated from the EOS table of Chabrier et al.(2019) and Chabrier \& Debras (2021). The first term of Equation (\ref{energy}) is the internal energy and the second term  is the gravitational potential energy. All the terms are calculated at the radiative-convective boundary.

\subsection{Boundary conditions}

In our fiducial case, the planet core mass $M_{core}$ is set to 5 $\mathrm{M_\oplus}$ with constant density $\rho_{core} = 7 \ g \ {cm}^{-3}$, same as Lee et al.(2014). The corresponding core radius is 1.6 $\mathrm{R_\oplus}$. The outer radius is determined by the minimum value of the Bondi radius and Hill radius:
\begin{equation}
    R_{out} = \min(R_\mathrm{B}, R_\mathrm{H})
\end{equation}
where
\begin{equation}
  R_\mathrm{H} = \left( \frac{M_p}{M_*}   \right)^{1/3} a
  \approx 40 R_\oplus \left( \frac{M_p}{5 M_\oplus}\right) ^{1/3} a_{0.1}  
\end{equation}
\begin{equation}
  R_\mathrm{B} =\frac{GM_p}{c_{s}^2} \approx 90 {R_\oplus}\left( \frac{M_p}{5M_\oplus} \right)T_3^{-1}
\end{equation}
where $a_{0.1}=a/0.1 AU$, $T_3 = T/1000K$.

The temperature and density of the outer boundary are set to the disk density and temperature at the planet location. The disk structure is from the minimum-mass extrasolar nebula model of Chiang \& Laughlin (2013):  

\begin{equation}
    T_\mathrm{d}=   10^3 \ a_{0.1}^{-3/7} K 
\end{equation}

\begin{equation}
    \rho_\mathrm{d}= 6 \times 10^{-6} a_{0.1}^{-2.9} g \ {cm}^{-3}
\end{equation}
We set $a= 0.1 AU$ as our fiducial case. Correspondingly, the disk temperature and density are 1000 K and $6 \times 10^{-6} \ \mathrm{g \ {cm}^{-3}}$, respectively.

\subsection{Calculation procedure}

The main goal of our model is to study the planetary interior structure and cooling contraction considering ohmic heating during the slow gas-accretion phase. Planetary interior structure and the corresponding ohmic heating profile are calculated self-consistently for a given set of boundary conditions and the parameters of the ohmic heating model. If the ohmic heating profile matches the interior thermal structure, the iteration and integration of our procedure are completed. The procedure is performed as follows:

(a) Choose a set of initial conditions and boundary conditions. The planetary thermal structure model is as follows: the planetary core mass ($M_\mathrm{c}$) and envelope mass ($M_\mathrm{envi}$); the outer boundary ($R_\mathrm{out},M_\mathrm{out}=M_\mathrm{c}+M_\mathrm{envi},T_\mathrm{out}=T_\mathrm{d},\rho_\mathrm{out}=\rho_\mathrm{d}$); the inner boundary ($R_\mathrm{in}=R_\mathrm{c},M_\mathrm{in}=M_\mathrm{c}$). The ohmic heating model is as follows: magnetic field $\mathbf{B}$, wind velocity $\mathbf{v}$ and wind zone depth $d_\mathrm{w}$; the outer boundary ($J_\mathrm{r} = 0$ at $R_\mathrm{out}$); the inner boundary ($J_\mathrm{r} = 0$ at $R_\mathrm{in}$).
 
(b) Assume or guess a total luminosity, $L=L_\mathrm{cool} + P_\mathrm{oh}$. Numerically solve the ODE Equations (7), (8), (10), (11)-(13) and (18) from $R_\mathrm{out}$ inward to $R_\mathrm{in}$. The Ohmic heating profile [ Equations (7),(8) and (10)] is iterated until it matches the given boundary conditions of ohmic heating model.

(c) Check the consistency of the envelope mass ($M_\mathrm{envf}$) obtained from step (b) and the initial setting mass ($M_\mathrm{envi}$) in step (a). If the tolerance between $M_\mathrm{envf}$ and $M_\mathrm{envi}$ is less than $10 ^{-5}$, the procedure is finished and the desired planetary thermal structure is found. If not, return to step (b).

(d) Increase the envelope mass and repeat steps (a)-(c) to obtain a series of thermal structures. Employ Equation (19) to calculate the cooling time.

\section{Results}\label{resul}

In this section, we present our numerical results for the planetary envelope thermal structure and cooling contraction taking into account the ohmic dissipation. For the sake of simplicity, we adopt the Ohmic heating model from Wu \& Lithwick (2013a) whose model assumes a dipolar magnetic field and a constant wind velocity in the wind zone.

The parameters of ohmic heating model are primarily related to the magnetic field acting on the planet, wind velocity profile, and wind zone depth. 
For the studies of hot Jupiters inflation, the wind velocity is estimated from the atmosphere's thermal dynamics because of the tidal locking of close-in planet (e.g., Showman et al. 2010; Menou 2012), on the order of 1 $\mathrm{km \ s^{-1}}$. The strength of the dipole magnetic field on the planetary surface is  commonly adopted on the order of 1 G. The bottom of the wind zone is normally set at the location of pressure with about 10 $bar$ (e.g.,Batygin \& Stevenson 2010; Wu \& Lithwick 2013a; Pu \& Valencia 2017) or estimated with pressure scale height $H_\mathrm{P}$ (e.g., Huang \& Cumming 2012; Menou 2012).

Although the exact values of these parameters are unknown for our case, we can get some references from numerical simulations and observational results. 
The atmospheric recycling flow can penetrate to the inner part of the planetary envelope around 0.3 $R_{out}$, (e.g., Lambrechts \& Lega 2017; Kurokawa \& Tanigawa 2018; Zhu et al. 2021). As the planetary mass increases, the azimuthal velocity of atmospheric flow increases close to the Keplerian velocity (e.g., Ormel et al.2015a, 2015b; B{\'e}thune \& Rafikov 2019a). Moreover, the planetary envelope can be rotational support during the accretion phase (e.g., Ormel et al.2015a; B{\'e}thune \& Rafikov 2019a, 2019b). Based on the results of these studies, we set the wind zone depth (from planetary surface to the bottom of wind zone) $d_w = 0.7 R_\mathrm{out}$ as our fiducial case. For a 5 $M_\oplus$ core locating at 0.1 AU, the Keplerian velocity at the Hill radius is about $10^5 \  \mathrm{ cm \ s^{-1}}$ and at the surface of planetary core is about $v_\mathrm{k} \simeq 10^6 \ \mathrm{cm \ s^{-1}}$. Since we do not know the rotational velocity of the planetary core, we set the fiducial wind velocity $v_\mathrm{w}=0.01 \ v_\mathrm{k}$, $10^4 \ \mathrm{cm \ s^{-1}}$.

The magnetic field of Earth-like planets can be generated by the movement of liquid iron within their cores (e.g., Cuartas-Restrepo 2018). Numerical simulations show that the surface magnetic field strength of super-Earths is around 1 G (Driscoll \& Olson 2011),  depending mainly on the iron-core mass fractions. Recent ultraviolet observations of HAT-P-11 show that this Neptune size planet (about 24 $M_\oplus$) should pose equatorial magnetic field strength with 1-5 G (Ben-Jaffel et al. 2022). Moreover, close-in protoplanets may suffer from the magnetic interaction from their pre-mainsequence hosts because the 
observed surface magnetic fields of T Tauri stars can be on the order of 1000 G (e.g., Johns-Krull 2007; Yang et al. 2008). For our our fiducial case, the surface  magnetic dipole field strength (the magnetic field strength $B$, hereafter) is set to $B = 1 G$.
Furthermore, we explore the influence of parameter space of ohmic heating model on the planetary cooling contraction. In our calculations, we consider wind zone depth $d_\mathrm{w}$ from 0.5 $R_\mathrm{out}$ (shallow) to 0.9 $R_\mathrm{out}$(deep). The magnetic field strength B spans from 0.1 G up to 10 G. The wind velocity is in a range from $10^2$ to $10^5 \ \mathrm{cm \ s^{-1}}$.

Examples of planetary envelope thermal structures with the external energy source from ohmic heating are presented in Section \ref{interior}. In Section \ref{tcool}, we show the variations of planetary luminosity and cooling time with increasing envelope mass. Parameter space is then explored to study the effect of ohmic dissipation on the planetary cooling contraction in Section \ref{para}. We also explore the influence of ohmic dissipation on the envelope evolution of protoplanets with dust opacity at the end of Section \ref{para}.

\begin{figure*}[h!]
\centering
\includegraphics[scale=0.5]{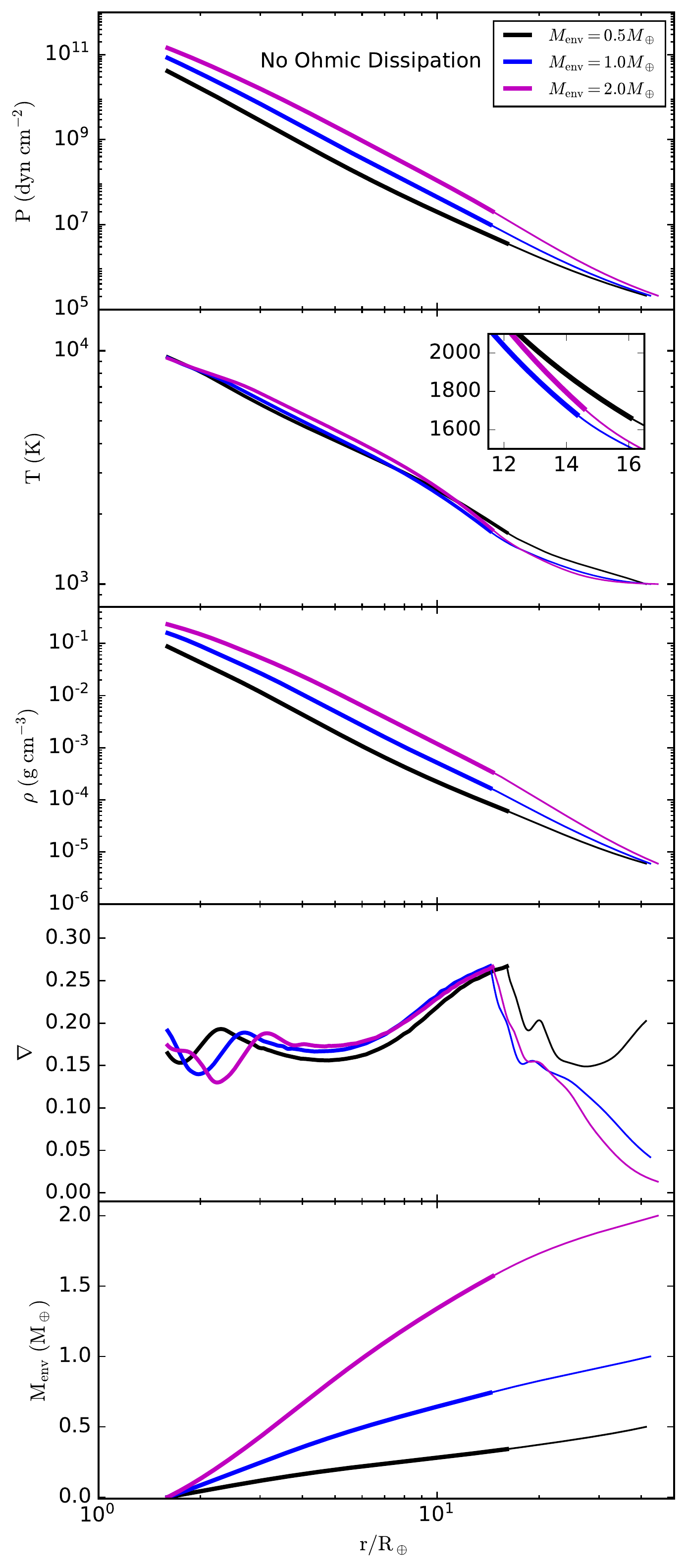}
\includegraphics[scale=0.5]{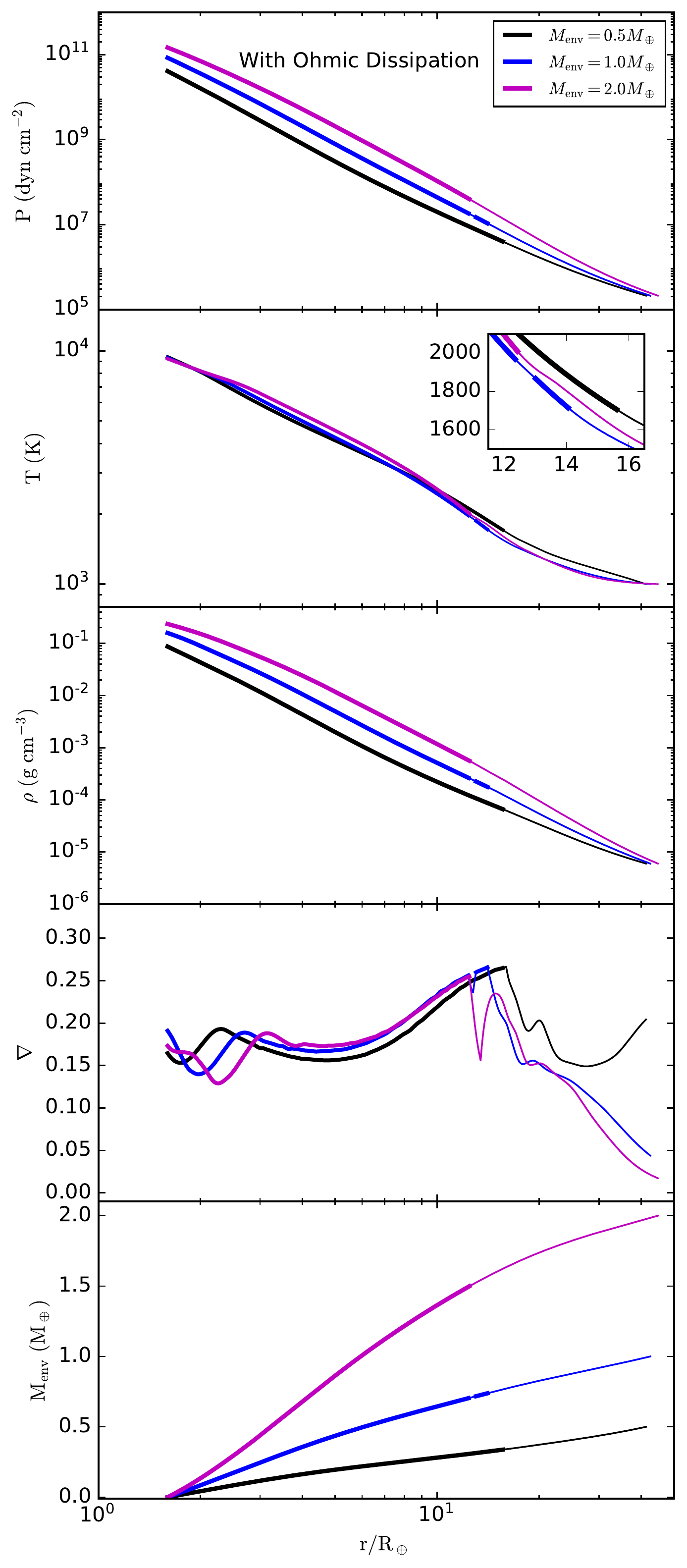}
\caption{Thermal structures of a planet with 5 $M_\oplus$ core mass during envelope growth. The planet locates at 0.1 AU. Radial profiles of three different envelope mass (0.5 $M_\oplus$, 1.0 $M_\oplus$, 2.0 $M_\oplus$) are shown. From top to bottom, the panels show the pressure profiles, temperature profiles, density profiles, temperature gradient profiles and envelope mass profiles, respectively. Thick lines denote the convective zone and the thin lines indicate the radiative layer. Left-hand panels present the structures without ohmic dissipation while the right-hand panels display the structures with ohmic dissipation. The parameters of ohmic dissipation model are our fiducial case (wind zone depth $d_w = 0.7 R_\mathrm{out}$, wind velocity $v_\mathrm{w}= \ 10^4 \ \mathrm{cm \ s^{-1}}$, magnetic field strength $B$= 1 G). The radiative-convective boundaries are pushed into the deeper part of the planet when ohmic dissipation is taken into account, the details are shown in the panels of temperature profiles.
The detailed model parameters are shown in Table \ref{tab1}. }
\label{fig:int}
\end{figure*}

\subsection{Planetary thermal structure} \label{interior}

\begin{figure}[h!]
\centering
\includegraphics[scale=0.55]{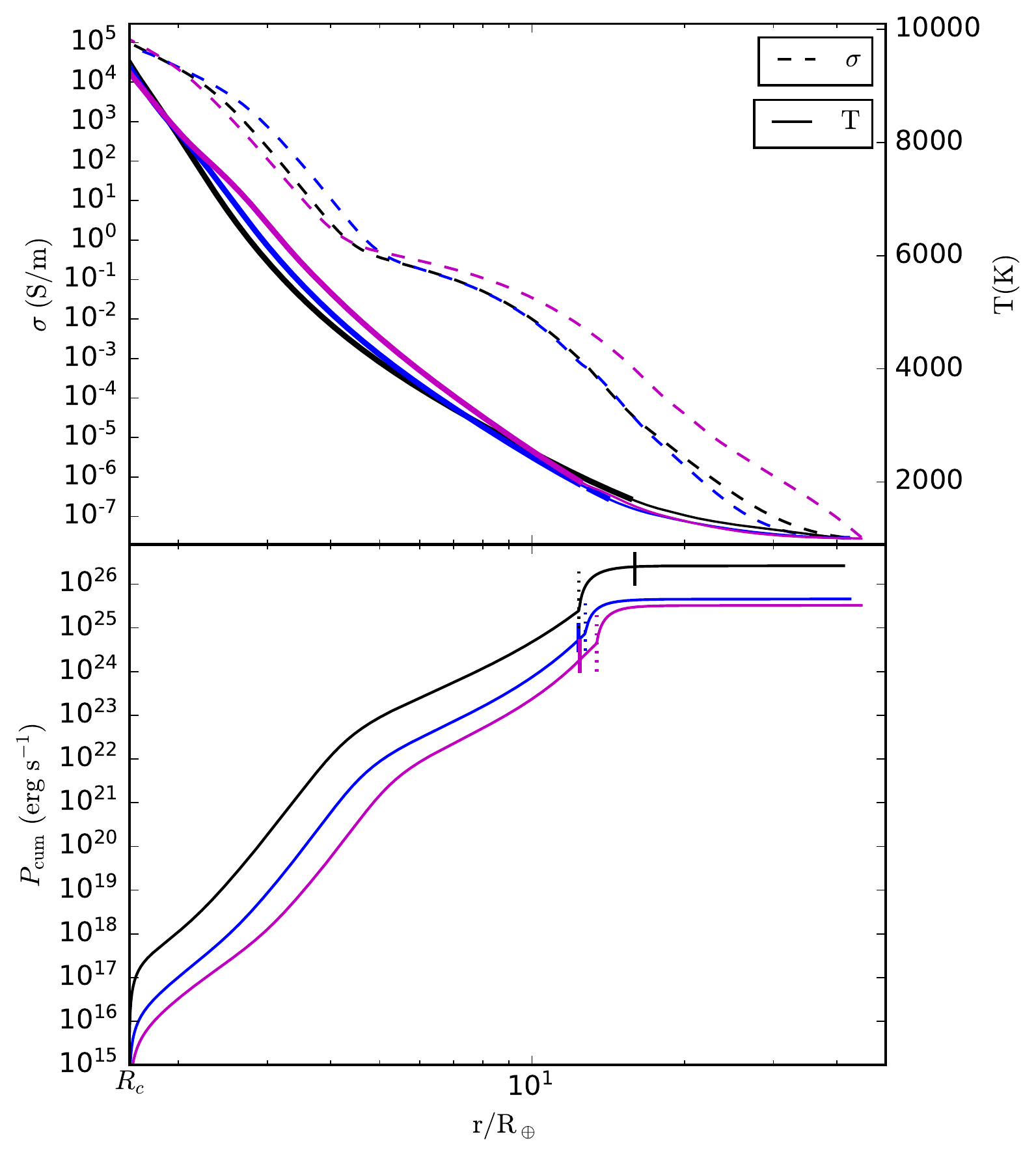}
\caption{The radial profiles of temperature (upper panel, solid lines), conductivity (upper panel, dashed lines) and cumulative ohmic heating power (bottom panel), corresponding to the thermal structures with ohmic dissipation in Figure \ref{fig:int}. The vertical lines in the bottom panel denote the radiative-convective boundaries. The vertical-dotted lines in the bottom panel represent the bottom of the wind zone.}
\label{fig:ps-int}
\end{figure}

Radial profiles of a planet with 5 $M_\oplus$ core at 0.1 AU during envelope growth are shown in Figure \ref{fig:int}. The left-hand panels show the interior structures without Ohmic dissipation while the right-hand panels with ohmic dissipation. The parameters of ohmic heating model here are our fiducial case ($d_w = 0.7 R_\mathrm{out}$, $v_\mathrm{w}= \ 10^4 \ \mathrm{cm \ s^{-1}}$, B= 1 G). The radiative-convective boundary is an essential factor to describe the planetary thermal structure and cooling contraction. We can see that the radiative-convective boundaries of these envelopes move inward when the effect of ohmic dissipation is considered (see the details of the inner plots of the temperature profiles and Table \ref{tab1}). Although the ohmic heating power contributes a large portion of energy ($\sim 65 \%$) to planetary luminosity for the envelope mass with 0.5 $M_\oplus$ (Table \ref{tab1}), most of the power ($\sim$ \ 95\%) is directly deposited into the convective zone (see the ohmic heating power profile in Figure \ref{fig:ps-int}). The small fraction of ohmic heating power contained in the outer envelope has a minor effect on the thermal structure (see the discussion below). As a result, the radiative-convective boundary is slight pushed inward (see the parameters at the radiative-convective boundary in Table \ref{tab1}). As the envelope mass grows, multiple layers appeare in the envelope (with 1 $M_\oplus$), which means that a radiative zone is formed between two adjacent convective layers, similar to the sandwiched structure mentioned in Lee et al.(2014). Compared to the low envelope mass (0.5 $M_\oplus$) at the initial stage of gas accretion, the radiative-convective boundaries are obviously pushed inward  for envelopes with high mass (1, 2 $M_\oplus$).

The overall properties of the ohmic power profile is that the most of the ohmic heating energy is concentrated into the wind zone and the energy deposited below the wind zone orders of magnitude smaller than that in the wind zone, as shown in the bottom panel of Figure \ref{fig:ps-int} (e.g., Huang \& Cumming 2012; Wu \& Lithwick 2013a). From the upper panel of Figure \ref{fig:ps-int}, we can see that the conductivity increases rapidly as temperature increases. The conductivity several orders of magnitude higher in deep part of planetary envelope than that in the upper envelope, leading to a sharp reduction of ohmic heating power (Equation \ref{eq:dpdr}). The vertical solid and dotted lines in Figure \ref{fig:ps-int} denote the radiative-convective boundary and the bottom of the wind zone, respectively. The effect of ohmic dissipation on the planetary thermal structure depends on the distribution of Ohmic heating power in the envelope. For the envelope with 0.5 $M_\oplus$, the bottom of wind zone is below the radiative-convective boundary and most of the ohmic heating energy ($\sim$ \ 95\%) is deposited into the convective zone to replace the cooling luminosity. However, as mentioned earlier, the energy contained in the upper envelope is not enough to efficiently change the structure of the envelope. The case for the envelope with 1 $M_\oplus$ (2 $M_\oplus$) is converse. The bottom of the wind zone is close to (above) the radiative-convective boundary and about $90 \% $ ($95 \% $) of ohmic heating power is concentrated into the outer envelope. The thermal structure of the envelope is changed obviously (Figure \ref{fig:int} and Table \ref{tab1}).
When the envelope mass increases to 2 $M_\oplus$, about 75 \% of planetary luminosity is from the Ohmic heating power (Table \ref{tab1}). The effect of ohmic dissipation becomes important and planetary cooling contraction starts to slow down significantly, as discussed in next section.

\begin{deluxetable*}{lcccccccccc}
\tablenum{1}
\tablewidth{1pt} 
\tablecaption{Parameters of thermal structures in Figure 1  \label{tab1}}
\tablehead{
\colhead{$M_\mathrm{env}$} & \colhead{Ohmic Dissipation} &  \colhead{L} & \colhead{$P_\mathrm{oh}$} & \colhead{$L_\mathrm{cool}$} &\colhead{$P_\mathrm{oh}/L$} & \colhead{$R_\mathrm{rcb}$} & \colhead{$M_\mathrm{rcb}$} & \colhead{$P_\mathrm{rcb}$} & \colhead{$T_\mathrm{rcb}$} &\colhead{$\rho_\mathrm{rcb}$}
\\
\colhead{($M_\oplus$)} & \colhead{} & \colhead{($\mathrm{erg \ s^{-1}}$)} & \colhead{($\mathrm{erg \ s^{-1}}$)} & \colhead{($\mathrm{erg \ s^{-1}}$)} & & \colhead{($R_\oplus$)} & \colhead{($M_\oplus$)} & \colhead{($\mathrm{bar}$)} & \colhead{$\mathrm{(K)}$} & \colhead{$\mathrm{(g \ {cm}^{-3})}$}
} 
\startdata
0.5 & No & $4.011\times 10^{26}$ &  & $4.011 \times 10^{26}$ & & 16.048 & 5.341 & 3.598 & 1662.6 & $6.134 \times 10^{-5}$  \\
0.5 & Yes & $4.046\times 10^{26}$ & $2.633 \times 10^{26}$ & $1.413 \times 10^{26}$ &  0.651 & 15.965 & 5.341 & 3.661 & 1669.8 & $6.214 \times 10^{-5}$ \\
1.0 & No & $9.090\times 10^{25}$ & & $9.090 \times 10^{25}$ &  & 14.330 & 5.742 & 9.944 & 1680.0 & $1.677 \times 10^{-4}$ \\
1.0 & Yes & $9.486\times 10^{25}$  & $4.571 \times 10^{25}$ & $4.915 \times 10^{25}$ & 0.482 & 12.355 & 5.704 & 18.339 & 1968.0 & $2.638 \times 10^{-4}$ \\
2.0 & No & $3.317\times 10^{25}$ & & $3.317 \times 10^{25}$ &  & 14.556 & 6.570 & 20.463 & 1711.3 &  $3.384 \times 10^{-4}$ \\
2.0 & Yes & $4.348\times 10^{25}$ & $3.267 \times 10^{25}$ & $1.080 \times 10^{25}$  & 0.752 & 12.430 & 6.499 & 39.887 & 2009.0 & $5.613 \times 10^{-4}$ \\
\enddata
\tablecomments{$M_\mathrm{env}$ is the planetary envelope mass. $L$ is the total luminosity of the planet. $\mathrm{P_{oh}}$ and $\mathrm{L_{cool}}$ are the ohmic heating power and planetary cooling luminosity. The subscript $\mathrm{rcb}$ denotes the radiative-convective boundary. For the thermal structure with multiple layers (a radiative zone is sandwiched between two adjacent convective zones, as the structure of the envelope with 1 $M_\oplus$ in Figure \ref{fig:int}), the parameters ($R_\mathrm{rcb}$, $M_\mathrm{rcb}$, $P_\mathrm{rcb}$, $T_\mathrm{rcb}$, $\rho_\mathrm{rcb}$) are chosen at the inner most boundary.}
\end{deluxetable*}

\subsection{Cooling contraction with Ohmic dissipation}\label{tcool}

\begin{figure}[h!]
\centering
\includegraphics[scale=0.6]{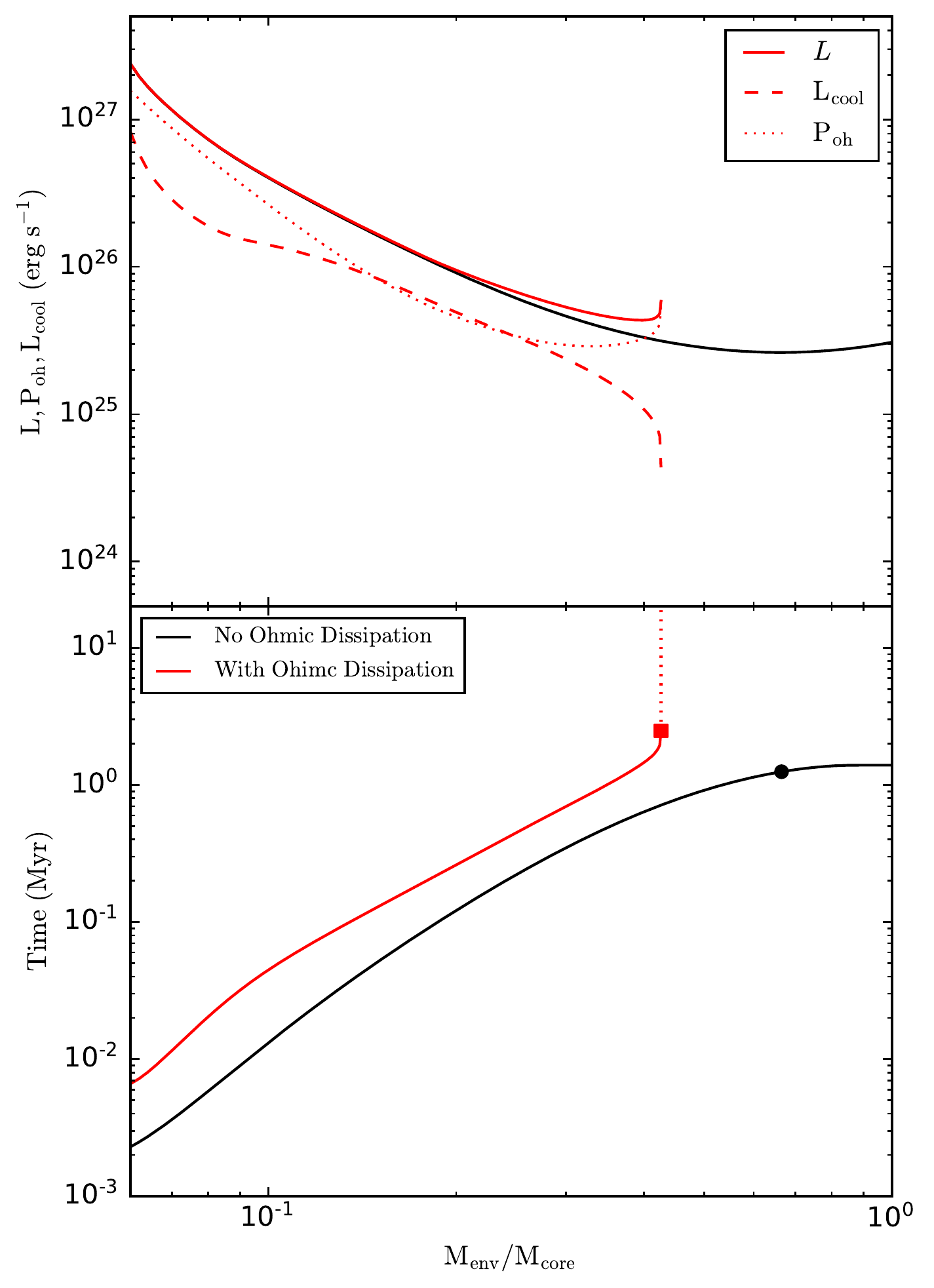}
\caption{The luminosity and evolutionary time vary with the envelope mass. The planetary core mass is 5 $M_\oplus$. The black line shows the case without Ohmic dissipation while the red lines describe the case with ohmic dissipation. The parameters of ohmic heating model that are used here are the same as in Figure \ref{fig:int}. Upper panel: solid lines represent the total luminosity of the planet. The dashed line and dotted line denote the cooling luminosity and the ohmic heating power, respectively. The circle and square symbols in the bottom panel designate the runaway time ($\mathrm{t_{run}}$) and the halting mass with omic dissipation ($M_\mathrm{halt}$), respectively. The 
vertical red-dotted line expresses that the contraction of planet is halted. Without ohmic dissipation, the planet starts runaway gas accretion at $\mathrm{t_{run}} \simeq$ 1 Myr. When the effect of ohmic dissipation is considered, gas accretion is halted at about 2 Myr and $M_\mathrm{halt} = 2.1 M_\oplus$, about 42 \% of planetary core mass. For further details, see Section 3.2.}
\label{fig:cool}
\end{figure}

Figure \ref{fig:cool} shows the variations of luminosity and cooling time with the envelope mass. Without ohmic dissipation (black lines in Figure \ref{fig:cool}), the planet starts runaway accretion at about 1 Myr\footnote{The time of runaway accretion ($\mathrm{t_{run}}$) is usually defined as the time when the minimum of the planetary luminosity is reached (e.g., Lee et al. 2014; Yu 2017).} (the black-solid circle in the bottom panel of Figure \ref{fig:cool}). When the effect of ohmic dissipation is taken into account, the planetary cooling is modulated by gravitational contraction and ohmic heating. With the envelope mass increasing, the luminosity (red solid line), cooling luminosity (red dashed line) and ohmic heating power (red dotted line) reduces. Until the planetary luminosity is totally dominated by ohmic heating power ($P_\mathrm{oh}/L > 90 \%$), the contraction of planet is significantly delayed. We stop the calculation when the ratio of ohmic heating power and total luminosity is around 0.99 ($P_\mathrm{oh}/L \simeq 99 \%$) and assume that gas accretion is halted. The halting mass $M_\mathrm{halt}$ is defined as the envelope mass when the ratio reaches 0.99. 
For our fiducial parameter set of the ohmic-heating model, gas accretion is halted at about 2 Myr and $M_\mathrm{halt} = 2.1 M_\oplus$, about 42 \% of planetary core mass, as shown in the bottom panel (red square) of Figure \ref{fig:cool}. We can see that the planet avoids runaway gas accretion and stops accreting gas when the effect of ohmic dissipation is coupled into the envelope accretion.

\subsection{Parameter study of Ohmic dissipation}\label{para}

In the previous section, only a set of ohmic heating model parameters are studied, which is not enough to investigate the effect of Ohmic dissipation on planetary cooling.
Moreover, we adopt a simplified ohmic-dissipation model which assumes that the magnetic filed strength and wind velocity are constant. This is not realistic during the envelope growth. The properties of magnetic field and wind zone should change simultaneously during envelope accretion. Thus, parameter study is needed. In this section, we explore the parameter space to study the effect of these parameters (i.e., the strength of magnetic, the wind zone depth, and wind velocity) on the planetary cooling. Envelope accretion with dust opacity is also briefly discussed.

\begin{figure}[h!]
\centering
\includegraphics[scale=0.7]{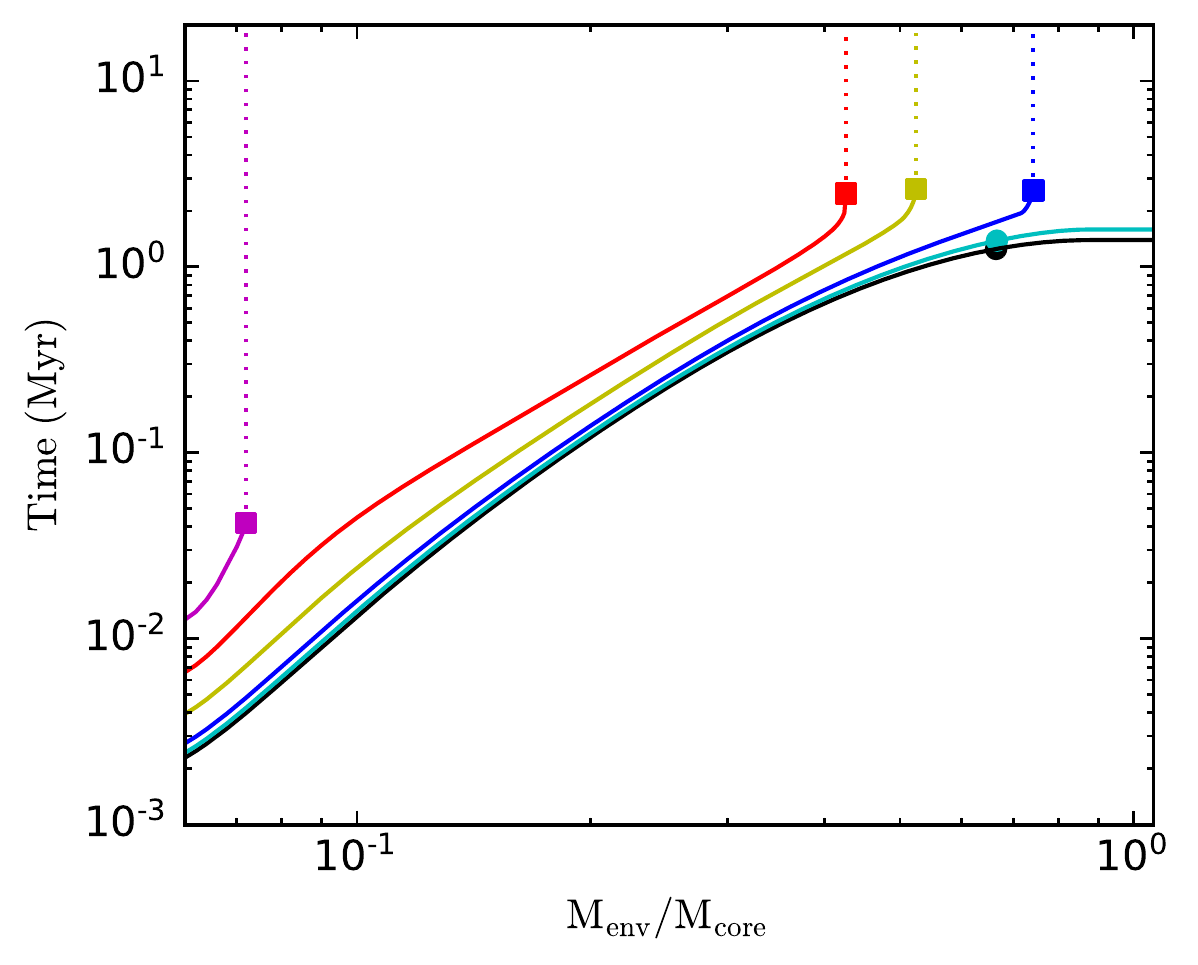}
\caption{Evolutionary time varies with envelope mass for different wind velocities. The magnetic strength is 1 G. The wind zone depth is 0.7 $R_\mathrm{out}$. From right-hand to left-hand, the different color lines shows cases for wind velocity with $v_\mathrm{w}=3 \times 10^3 \ \mathrm{cm \ s^{-1}}$ (cyan), $v_\mathrm{w}=5 \times 10^3 \ \mathrm{cm \ s^{-1}}$ (blue),$v_\mathrm{w}=8 \ \times 10^3 \ \mathrm{cm \ s^{-1}}$ (yellow), $v_\mathrm{w}=1 \times 10^4 \ \mathrm{cm \ s^{-1}}$ (red), $v_\mathrm{w}=1.12 \times 10^4 \ \mathrm{cm \ s^{-1}}$ (magenta), respectively. The black line denotes the case without ohmic dissipation.}
\label{fig:lvt}
\end{figure}

\begin{figure}[h!]
\centering
\includegraphics[scale=0.65]{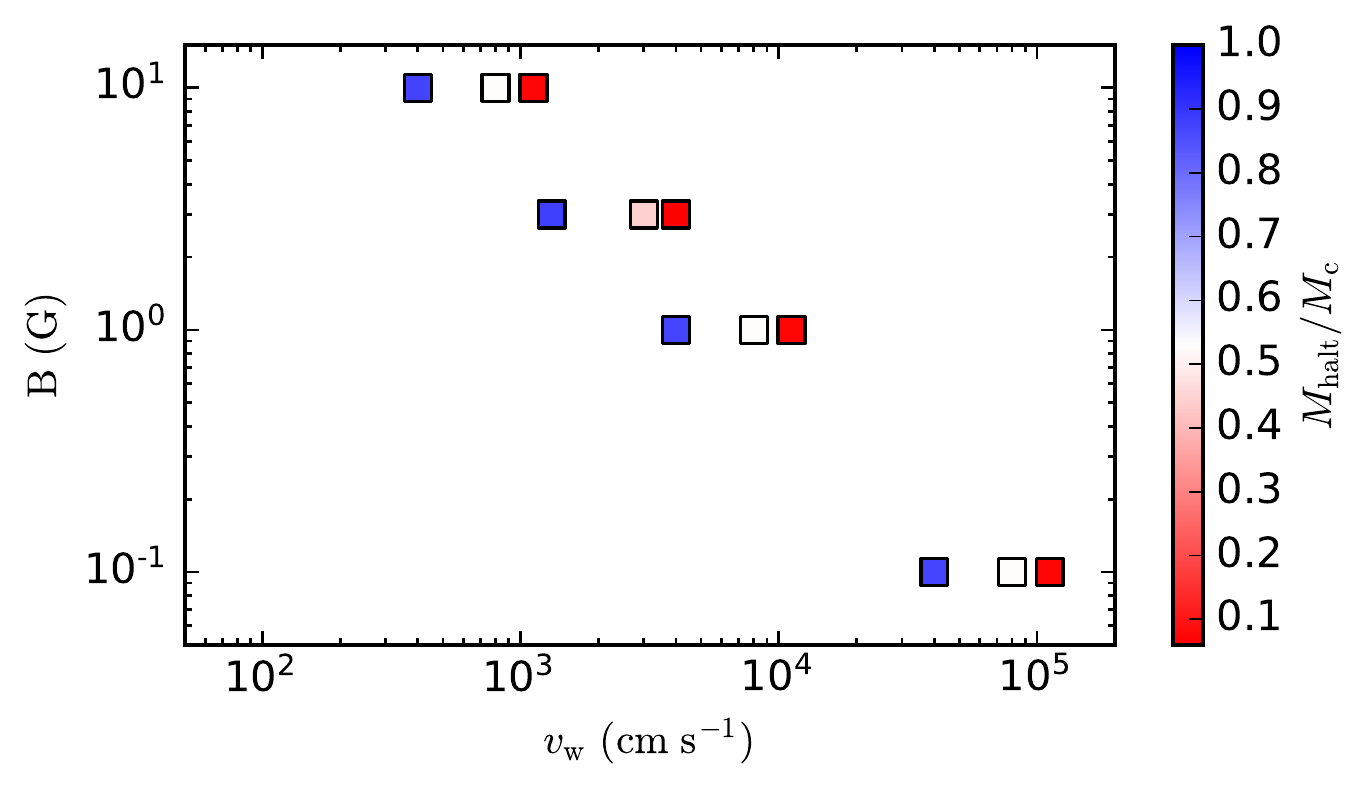}
\caption{The halting mass varies with the magnetic field strength and wind velocity. The wind zone depth is 0.7 $R_\mathrm{out}$. The planetary core here is 5 $M_\oplus$. The color bar denotes the ratio of halting mass and the core mass ($\mathrm{M_{halt}/M_c}$). The largest values of the ratio (blue squares) are less than 1 (around 0.9). For the parameters located on the left-hand side of blue squares, ohmic dissipation has a negligible effect on the planetary cooling and runaway gas accretion is unavoidable. When the parameters of ohmic heating model are located on the right side of the red squares,  gas cannot be accreted onto the planet due to the strong ohmic dissipation. }
\label{fig:bvm}
\end{figure}

\subsubsection{Wind velocity and magnetic field}

Figure \ref{fig:lvt} shows the cooling time with increasing envelope mass for different wind velocities. The magnetic field strength is 1 G and the wind zone depth 0.7 $R_\mathrm{out}$. From right-hand to left-hand, the wind velocity increases from $3 \times 10^3$ to $1.12 \times 10 ^4 \mathrm{cm \ s^{-1}}$. As wind velocity increases, the gas accretion is significantly delayed and the halting mass reduces. As the wind velocity reaches $1.12 \times 10^4 \mathrm{cm \ s^{-1}}$, the planet stops contraction and accretion, even when the envelope mass is just several percent (about 7\%) of the core mass (magenta line in Figure \ref{fig:lvt}). A super-Earth may be formed in this case.

\begin{figure}[h!]
\centering
\includegraphics[scale=0.65]{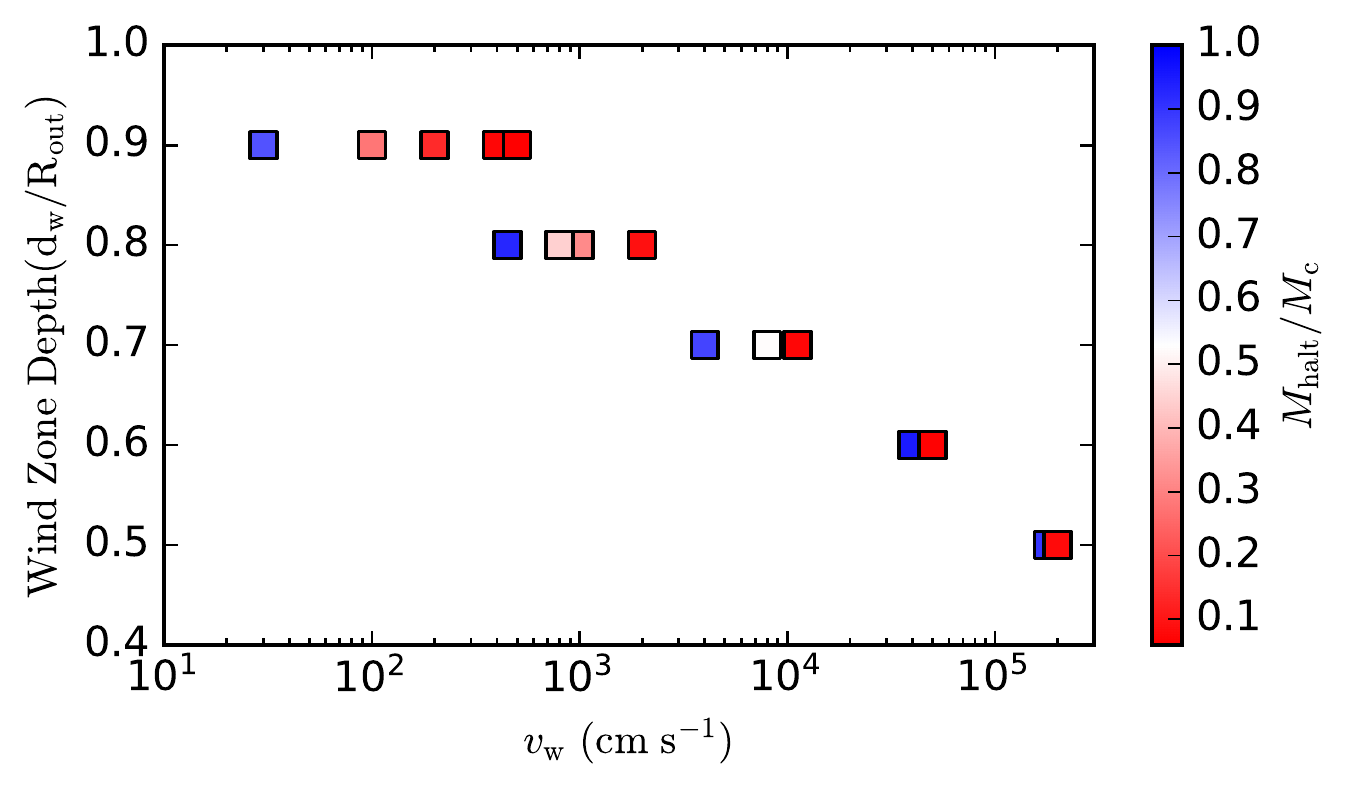}
\caption{Same as Figure \ref{fig:bvm}, but the halting mass varies with the wind zone depth and wind velocity. The magnetic field strength is 1 G.}
\label{fig:rvm}
\end{figure}

Figure \ref{fig:bvm} shows the halting mass with different values of magnetic field strength and wind velocity for the core mass with 5 $M_\oplus$ . The wind zone depth is 0.7 $R_\mathrm{out}$. Different colors denote the value of the ratio of the halting mass and the core mass, as the color bar shown in the figure. The blue squares in Figure \ref{fig:bvm} (or Figure \ref{fig:rvm}) represent the largest values of the ratio can be reached, generally around 0.9 (but less than 1). The effect of ohmic dissipation is negligible for the parameters located on the left-hand side of the blue squares. The planet undergoes runaway gas accretion in this situation. 
As the wind velocity and magnetic field strength increase, the halting mass reduces. The halting mass can be several percent of core mass for some parameter sets, as the red squares shown in Figure \ref{fig:bvm}\footnote{When the parameters are located on the right-hand side of red squares, the ohmic dissipation is so strong that gas cannot be accreted onto the planet.}. If the planet possess strong magnetic field, a relative small wind velocity can efficiently halt the gas accretion, and vice versa. The magnetic field acting on the protoplanets needs to be constrained by future observations.

Figure \ref{fig:bvm} also implies that the case of small ratio ($\mathrm{M_{halt}/M_c} <0.1$) that prefers to form super-Earths can only be reached in a specific parameter space according to our Ohmic dissipation model. Although the parameter space can be expanded if different values of wind zone depth are considered (as discussed in next the section), it possibly originates from the assumptions that the wind velocity and magnetic field are constant in our calculations. In real cases, there should be a wind velocity (magnetic field) profile in the envelope and the velocity (magnetic field) should evolve with the envelope mass growth. More reliable results may be obtained if these factors are taken into account. Thus, the importance of ohmic dissipation on the envelope accretion need to be explored with a more realistic model.

\subsubsection{The wind zone depth}

Note that most of the ohmic power concentrated into the wind zone (e.g., appendix A of Wu \& Lithwick 2013a), meaning that 
the wind zone depth is essential for the radial distribution of ohmic heating power. 
The bottom of the wind zone determines the efficiency of ohmic power deposited into the inner convective zone. The extent of planetary contraction prevented by ohmic dissipation depends on how much of the cooling luminosity can be replaced by ohmic heating power during envelope accretion. 
With numerical results from previous studies (e.g., Lambrechts \& Lega 2017; Kurokawa \& Tanigawa 2018; Zhu et al. 2021), we set $d_{\mathrm{w}} \ = 0.7 \ R_\mathrm{out}$ as our fiducial case. Here, we explore the dependence of halting mass on the wind zone depth, as shown in Figure \ref{fig:rvm}. The strength of magnetic field is 1 G. The wind zone depth is from 0.5 to 0.9 $R_\mathrm{out}$. AS the wind zone depth increases, the wind velocity required to stop accretion reduces remarkably.
We can see that the planetary gas accretion is efficiently delayed for a deep wind zone ($\mathrm{d_w} =  0.9 \ R_\mathrm{out}$) with a wind velocity only $\mathrm{v_w}= \  10^2 \ \mathrm{cm \ s ^{-1}}$, which is in the order of $10^{-4}$ Keplerian velocity at the surface of planetary core.

\begin{figure}[h!]
\centering
\includegraphics[scale=0.6]{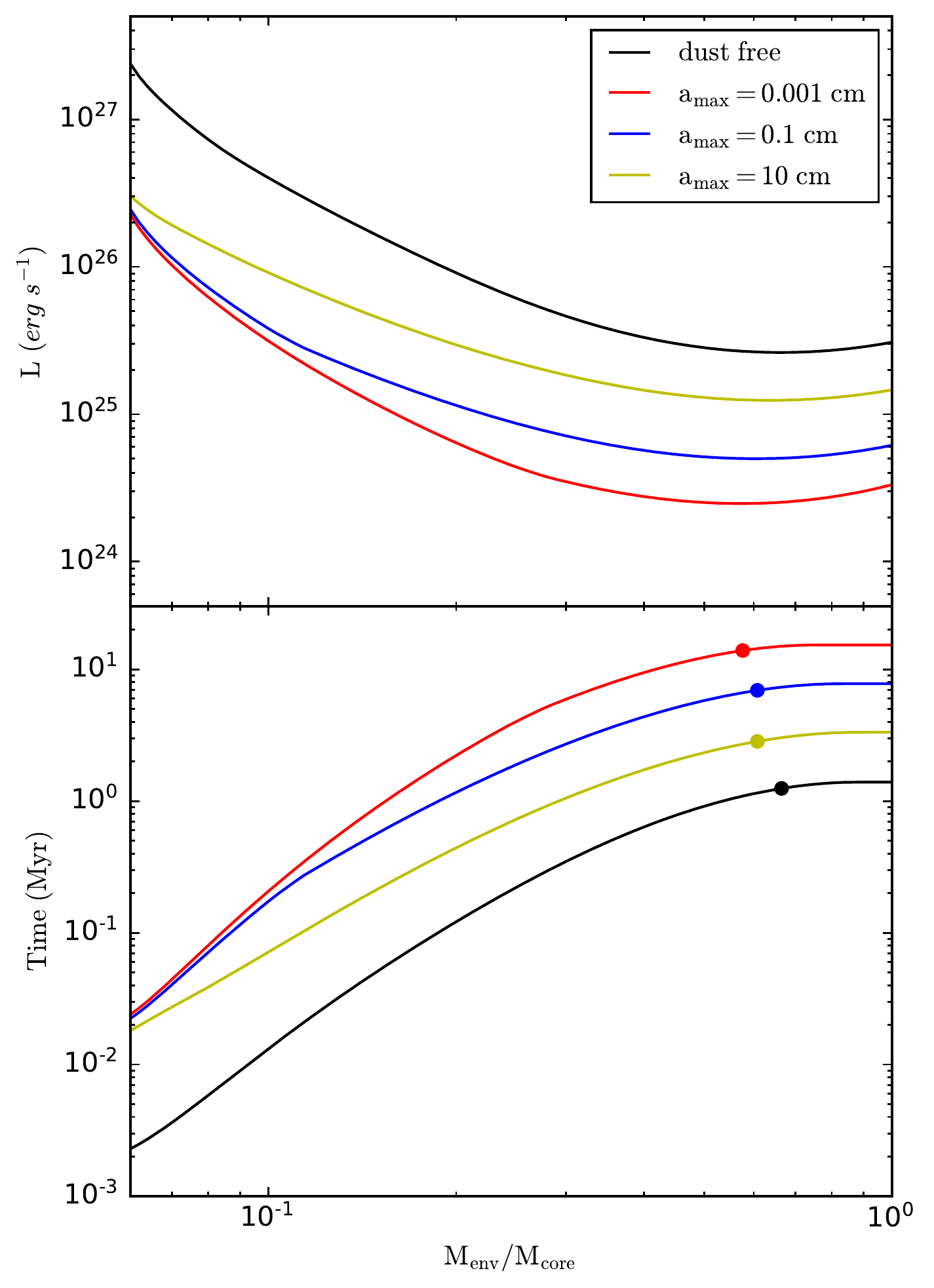}
\caption{The cooling luminosity and time with the increasing envelope mass when dust opacity is taken into account. The effect of ohmic dissipation is not included. $\mathrm{a_{max}}$ is the maximum particle size of dust grains.}
\label{fig:dust-lmt}
\end{figure}

\subsubsection{Dust opacity}\label{topa}

Opacity is crucial for the thermal structure of planetary envelope and dominates envelope contraction.  Compared with the dust-free opacity, the planetary cooling contraction is delayed by employing the dust opacity  (e.g., Lee et al. 2014; see also Figure \ref{fig:dust-lmt}). However, the opacity of planetary envelope during gas accretion is still unclear. Due to the efficient coagulation of dust grains in the envelope, a dust-free opacity may be favored (e.g., Mordasini et al. 2014; Ormel 2014).

We adopt the combined opacity from Zhu et al. (2021) to study the accretion process. Figure \ref{fig:dust-lmt} shows the variations of planetary luminosity and cooling time with different dust opacities\footnote{The effect of ohmic dissipation is not included here.}. For the situation of dust opacity with maximum dust grain size $a_\mathrm{max} = 0.001 \ \mathrm{cm}$, the runaway accretion time is about 10 Myr, which is similar to the result of the fiducial model from Lee et al. (2014).

\begin{figure}[h!]
\centering
\includegraphics[scale=0.7]{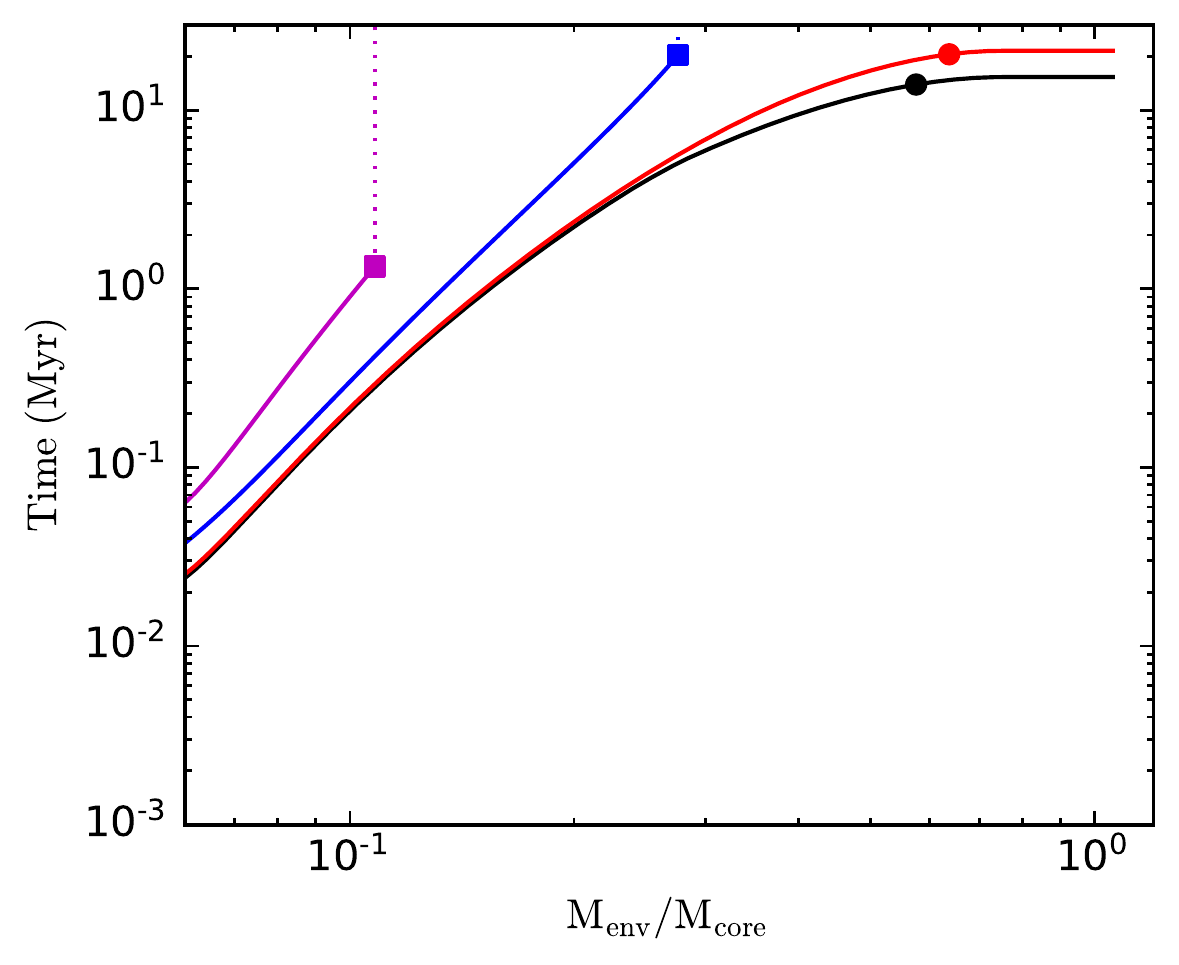}
\caption{The cooling time for the gas accretion of dusty envelope ($a_\mathrm{max} = 0.001 \ \mathrm{cm}$) with ohmic dissipation. The magnetic strength is 1 G. The wind zone depth is 0.7 $R_\mathrm{out}$. From right-hand to left-hand, the different color lines show cases for wind velocity with $v_\mathrm{w}=1 \times 10^3 \ \mathrm{cm \ s^{-1}}$ (red),$v_\mathrm{w}=3 \ \times 10^3 \ \mathrm{cm \ s^{-1}}$ (blue), $v_\mathrm{w}=5 \times 10^3 \ \mathrm{cm \ s^{-1}}$ (magenta), respectively. The black line denotes the case without ohmic dissipation.}
\label{fig:dust-lpm}
\end{figure}

Figure \ref{fig:dust-lpm} shows the cooling time with ohmic dissipation for the gas accretion of dusty envelope ($a_\mathrm{max} = 0.001 \ \mathrm{cm}$). The parameters of ohmic heating model are same as those of Figure \ref{fig:lvt}. We can see that the gas accretion of dusty envelope can be halted more effectively by ohmic dissipation relative to that of dust-free envelope. For example, the ratio of halting mass and planetary core mass is about 0.11 for the dusty envelope when the wind velocity $v_\mathrm{w} = 5 \times 10^3 \ \mathrm{cm \ s^{-1}}$, while the ratio is about 0.74 for the dust-free envelope (Figure \ref{fig:lvt}).

\section{Summary and Discussion} \label{concl}

The mechanism of ohmic dissipation has been widely used to study the radius inflation of close-in hot planets. For the close-in protoplanets, ohmic dissipation can also be induced in the planetary envelope as the disk gas is accreted onto the planetary surface and flows across the planetary intrinsic magnetic field or the field from its host star. In this work, we employ the ohmic-dissipation model from Wu \& Lithwick (2013a) to numerically investigate the effect of ohmic dissipation on the interior structure and cooling contraction of planetary envelopes during the slow accretion phase. We explore how the planetary cooling changes with different parameter sets of ohmic dissipation model.

The main findings are as follows: 

(1) With the ohmic-dissipation coupling into the planetary thermal structure, the cooling luminosity is replaced by the extra energy from ohmic heating deposited into the planetary envelope. As a result, the radiative-convective boundary of planetary envelope moves inward and envelope contraction slows down. Planetary envelope accretion is efficiently delayed and even halted when planetary luminosity is dominated by the ohmic heating power. 

(2) For some parameter sets of the ohmic heating model, the planetary envelope stops contraction and gas accretion is halted before  runaway accretion. The final envelope mass is only several percent of its core mass. Super-Earths may be formed in situ for these parameter sets of the ohmic heating model.

However, the ohmic-dissipation model and parameters we used have several limitations. We simply assume that the magnetic field and wind velocity are constant. Actually, the magnetic field and wind velocity should evolve as the envelope mass increases. The simple geometries of magnetic field and wind zone (i.e., wind zone depth and wind velocity profile) that are assumed in our calculations may not apply in real situations. We use approximate equations to calculate the conductivity, which may affect the ohmic heating profiles. 
In addition, the magnetic field acting on the protoplanets is not well constrained from observations. We simply assume a magnetic dipole field with surface strength from 0.1 to 10 G in our calculations. A more accurate ohmic-dissipation model with different boundary conditions and initial conditions are required to investigate the gas accretion of protoplanets in future work.
Although our model has these limitations, as a first step, our results show that ohmic dissipation is a potential mechanism to prevent planetary cooling contraction and promote the formation of super-Earths. Future observations of protoplanets may help to constrain the importance of ohmic dissipation on the super-Earth formation. 
 
\begin{acknowledgments}

We thank the anonymous referee for insightful comments and very useful suggestions to significantly improve our draft. We thank Dr. Liang Yin for useful discussions. This work has been supported by the National SKA Program of China (Grant No. 2022SKA0120101), Macau Science and Technology Development Fund (File No. 0001/2019/A1, No. 0051/2021/A1 and No. SKL-LPS(MUST)-2021-2023), the National Key R\&D Program of China (No. 2020YFC2201200) and the science research grants from the China Manned Space Project (No. CMS-CSST- 2021-B09 and CMS-CSST-2021-A10). C.Y. has been supported by the National Natural Science Foundation of China (grants 11373064, 11521303, 11733010, and 11873103), Yunnan National Science Foundation (grant Q9 2014HB048), and Yunnan Province (2017HC018).

\end{acknowledgments}

\appendix
\section{Ohmic heating Equation}\label{appexa}
Equations (7) and (8) can be changed as follows:
\begin{equation}
    {\Phi}_{2}'' + \left(\frac{d \ln \sigma}{d r} +\frac{2}{r}\right){\Phi}_{2}' - \frac{6}{r^2} {\Phi}_{2}- \left(\frac{d \ln \sigma}{d r} \frac{2\omega M_B}{3c r^2} + \frac{4\omega M_B}{c r^3}\right)=0, \ \  r_\mathrm{dep} \leq r < R_\mathrm{out}
\end{equation}
\begin{equation}
    {\Phi}_{2}'' + \left(\frac{d \ln \sigma}{d r} +\frac{2}{r}\right){\Phi}_{2}' - \frac{6}{r^2} {\Phi}_{2}=0,\ \  0<r<r_\mathrm{dep}\\
\end{equation}
where $\Phi$ is expressed with $\Phi=\Phi_2(r)P_2$ (Wu \& Lithwick 2013a), $P_2$ is the Legendre polynomial, $P_2 = \frac{1}{2} (3 \mathrm{cos}^2 \theta -1 )$. $r_\mathrm{dep}$ denotes the location of the bottom of the wind zone, $r_\mathrm{dep} = R_\mathrm{out} - d_\mathrm{w}$, where $d_\mathrm{w}$ is the wind zone depth. These two ordinary differential equations can be numerically solved.


\begin{references}

Ali-Dib, M., Cumming, A., \& Lin, D.~N.~C.\ 2020, \mnras, 494, 2440. doi:10.1093/mnras/staa914

Batalha, N.~M., Rowe, J.~F., Bryson, S.~T., et al.\ 2013, \apjs, 204, 24. doi:10.1088/0067-0049/204/2/24

Batygin, K. \& Stevenson, D.~J.\ 2010, \apjl, 714, L238. doi:10.1088/2041-8205/714/2/L238

Batygin, K., Stevenson, D.~J., \& Bodenheimer, P.~H.\ 2011, \apj, 738, 1. doi:10.1088/0004-637X/738/1/1

Ben-Jaffel, L., Ballester, G.~E., Garc{\'\i} Mu{\~n}oz, A., et al.\ 2022, Nature Astronomy, 6, 141. doi:10.1038/s41550-021-01505-x

B{\'e}thune, W. \& Rafikov, R.~R.\ 2019a, \mnras, 487, 2319. doi:10.1093/mnras/stz1427

B{\'e}thune, W. \& Rafikov, R.~R.\ 2019b, \mnras, 488, 2365. doi:10.1093/mnras/stz1870

Chabrier, G., Mazevet, S., \& Soubiran, F.\ 2019, \apj, 872, 51. doi:10.3847/1538-4357/aaf99f

Chabrier, G. \& Debras, F.\ 2021, \apj, 917, 4. doi:10.3847/1538-4357/abfc48

Chiang, E. \& Laughlin, G.\ 2013, \mnras, 431, 3444. doi:10.1093/mnras/stt424

Cuartas-Restrepo, P.\ 2018, Open Astronomy, 27, 183. doi:10.1515/astro-2018-0026

Dittmann, A.~J.\ 2021, \mnras, 508, 1842. doi:10.1093/mnras/stab2682

Driscoll, P. \& Olson, P.\ 2011, \icarus, 213, 12. doi:10.1016/j.icarus.2011.02.010

Freedman, R.~S., Lustig-Yaeger, J., Fortney, J.~J., et al.\ 2014, \apjs, 214, 25. doi:10.1088/0067-0049/214/2/25

Ginzburg, S. \& Sari, R.\ 2017, \mnras, 464, 3937. doi:10.1093/mnras/stw2637

Howard, A.~W., Marcy, G.~W., Johnson, J.~A., et al.\ 2010, Science, 330, 653. doi:10.1126/science.1194854

Huang, X. \& Cumming, A.\ 2012, \apj, 757, 47. doi:10.1088/0004-637X/757/1/47

Inamdar, N.~K. \& Schlichting, H.~E.\ 2015, \mnras, 448, 1751. doi:10.1093/mnras/stv030

Johns-Krull, C.~M.\ 2007, \apj, 664, 975. doi:10.1086/519017

Kurokawa, H. \& Tanigawa, T.\ 2018, \mnras, 479, 635. doi:10.1093/mnras/sty1498

Lee, E.~J., Chiang, E., \& Ormel, C.~W.\ 2014, \apj, 797, 95. doi:10.1088/0004-637X/797/2/95

Lee, E.~J. \& Chiang, E.\ 2015, \apj, 811, 41. doi:10.1088/0004-637X/811/1/41

Lee, E.~J., Chiang, E., \& Ferguson, J.~W.\ 2018, \mnras, 476, 2199. doi:10.1093/mnras/sty389

Liu, J., Goldreich, P.~M., \& Stevenson, D.~J.\ 2008, \icarus, 196, 653. doi:10.1016/j.icarus.2007.11.036

Liu, S.-F., Hori, Y., Lin, D.~N.~C., et al.\ 2015, \apj, 812, 164. doi:10.1088/0004-637X/812/2/164

Laine, R.~O., Lin, D.~N.~C., \& Dong, S.\ 2008, \apj, 685, 521. doi:10.1086/589177

Laine, R.~O. \& Lin, D.~N.~C.\ 2012, \apj, 745, 2. doi:10.1088/0004-637X/745/1/2

Lambrechts, M. \& Lega, E.\ 2017, \aap, 606, A146. doi:10.1051/0004-6361/201731014

Lopez, E.~D. \& Fortney, J.~J.\ 2014, \apj, 792, 1. doi:10.1088/0004-637X/792/1/1

Menou, K.\ 2012, \apj, 745, 138. doi:10.1088/0004-637X/745/2/138

Mordasini, C.\ 2014, \aap, 572, A118. doi:10.1051/0004-6361/201423702

Ormel, C.~W.\ 2014, \apjl, 789, L18. doi:10.1088/2041-8205/789/1/L18

Ormel, C.~W., Kuiper, R., \& Shi, J.-M.\ 2015a, \mnras, 446, 1026. doi:10.1093/mnras/stu2101

Ormel, C.~W., Shi, J.-M., \& Kuiper, R.\ 2015b, \mnras, 447, 3512. doi:10.1093/mnras/stu2704

Piso, A.-M.~A. \& Youdin, A.~N.\ 2014, \apj, 786, 21. doi:10.1088/0004-637X/786/1/21

Perna, R., Menou, K., \& Rauscher, E.\ 2010, \apj, 724, 313. doi:10.1088/0004-637X/724/1/313

Petigura, E.~A., Marcy, G.~W., \& Howard, A.~W.\ 2013, \apj, 770, 69. doi:10.1088/0004-637X/770/1/69

Pu, B. \& Valencia, D.\ 2017, \apj, 846, 47. doi:10.3847/1538-4357/aa826f

Rogers, L.~A.\ 2015, \apj, 801, 41. doi:10.1088/0004-637X/801/1/41

Showman, A.~P., Cho, J.~Y.-K., \& Menou, K.\ 2010, Exoplanets, 471

Weiss, L.~M. \& Marcy, G.~W.\ 2014, \apjl, 783, L6. doi:10.1088/2041-8205/783/1/L6

Wu, Y. \& Lithwick, Y.\ 2013a, \apj, 763, 13. doi:10.1088/0004-637X/763/1/13

Wu, Y. \& Lithwick, Y.\ 2013b, \apj, 772, 74. doi:10.1088/0004-637X/772/1/74

Wolfgang, A. \& Lopez, E.\ 2015, \apj, 806, 183. doi:10.1088/0004-637X/806/2/183

Yang, H., Johns-Krull, C.~M., \& Valenti, J.~A.\ 2008, \aj, 136, 2286. doi:10.1088/0004-6256/136/6/2286

Yu, C.\ 2017, \apj, 850, 198. doi:10.3847/1538-4357/aa9849

Zhong, W. \& Yu, C.\ 2021, \apj, 922, 215. doi:10.3847/1538-4357/ac2cc5

Zhu, W., Petrovich, C., Wu, Y., et al.\ 2018, \apj, 860, 101. doi:10.3847/1538-4357/aac6d5

Zhu, Z., Jiang, Y.-F., Baehr, H., et al.\ 2021, \mnras, 508, 453. doi:10.1093/mnras/stab2517

\end{references}
\end{document}